\documentstyle[12pt]{article}
\normalsize
\def\simless{\mathbin{\lower 3pt\hbox{$\rlap{\raise 5pt\hbox{$\char'074$}}
\mathchar"7218$}}}
\def\simgreat{\mathbin{\lower 3pt\hbox{$\rlap{\raise 5pt \hbox{$\char'076$}}
\mathchar"7218$}}}
\catcode`@=11

\def\beqra{\begin{eqnarray}} \def\eeqra{\end{eqnarray}}
\def\beq{\begin{equation}}      \def\eeq{\end{equation}}

\def\fo{\hbox{{1}\kern-.25em\hbox{l}}}

\def\ch{\@startsection{section}{1}{\z@}{-3ex plus-1ex minus-.2ex}%
        {2ex plus.2ex}{\large\sc}}

\def\; \lapp \;{\raisebox{-.4ex}{\rlap{$\sim$}} \raisebox{.4ex}{$<$}}

\def\con{\ifmmode \hbox{\bf*} \else{\bf*}\fi}   
\def\scon{\ifmmode \hbox{\footnotesize\rm\bf*} \else{\footnotesize\rm\bf*}\fi}

\def\0#1{\relax\ifmmode\mathaccent"7017{#1}
        \else\accent23#1\relax\fi}              

\def\eslash{\not{\hbox{\kern-2pt $E$}}}

\catcode`@=12

\begin{document}
\hoffset=0.4cm
\voffset=-1truecm
\normalsize
\arabic{page}
\def\ni{{\bar {N_i}}}    \def\nj{{\bar {N_j}}}   \def\n3{{\bar {N_3}}}
\def\li{\lambda_i}    \def\lj{\lambda_j}   \def\l3{\lambda_3}
\def\hn{h^\nu}       \def\hnij{h^\nu_{ij}}
\baselineskip=5pt
\begin{flushright}
DFPD 93/TH/06
\end{flushright}
\begin{flushright}
UTS-DFT-93-1
\end{flushright}
\begin{flushright}
hep-ph 9302207
\end{flushright}
\vspace{24pt}
\begin{center}

{\Large {\bf Spontaneous CP Violation in the Minimal Supersymmetric
Standard Model at Finite Temperature}}
\end{center}
\vspace{24pt}
\centerline{\large Denis Comelli}
\vskip 0.2 cm
\centerline{\it Dipartimento di Fisica Teorica Universit\`a di Trieste,}
\centerline{\it Strada Costiera 11, 34014 Miramare, Trieste, Italy}
\vskip 0.1 cm
\centerline{\it and}
\vskip 0.1 cm
\centerline{\it Istituto Nazionale di Fisica Nucleare,}
\centerline{\it Sezione di Trieste, 34014 Trieste, Italy}
\vskip 0.2 cm
\centerline{\it and}
\vskip 0.2 cm
\centerline{\large Massimo Pietroni}
\vskip 0.2 cm
\centerline{\it Dipartimento di Fisica Universit\`a di Padova,}
\centerline{\it Via Marzolo 8, 35100 Padua, Italy.}
\vskip 0.1 cm
\centerline{\it and}
\vskip 0.1 cm
\centerline{\it Istituto Nazionale di Fisica Nucleare,}
\centerline{\it Sezione di Padova, 35100 Padua, Italy.}

\baselineskip=30pt
\centerline{\bf {\large Abstract.}}
We show that in the Minimal Supersymmetric  Standard Model one--loop effects
at finite temperature may lead to a spontaneous breaking of
CP invariance in the scalar sector. Requiring that the breaking takes
place at the critical temperature for the electroweak phase transition,
we find that the parameters space is compatible with a mass of the Higgs
pseudoscalar in agreement with the present experimental lower bounds.
Possible implications for baryogenesis are discussed.

\newpage
\pagenumbering{arabic}
The possibility of a spontaneous breakdown of the CP symmetry in the scalar
sector of the Minimal Supersymmetric Standard Model (MSSM), has been recently
investigated \cite{Maekawa,Pomarol}. It is well known that, as long as
supersymmetry (SUSY) is exact, CP is conserved in the scalar sector of
the MSSM.
As a consequence, the only CP--violating effects may arise from the soft
SUSY--breaking terms: the scalar masses, the gaugino masses, and the trilinear
interactions.  If one allows these terms to be complex, then
there are two
new physical phases in the MSSM which are not present in the Standard
Model \cite{dugan}. These phases do not appear in the tree--level Higgs
potential, but occur in interactions involving the super--partners
of the ordinary particles, giving new contributions to the CP--odd
observables $\varepsilon$, $\varepsilon'$, and the electric dipole moment of
the neutron \cite{dugan}.
If one assumes that all the soft masses and couplings
are real, then no new CP violation appears at the tree--level.

At the one--loop level, the  contributions
of graphs with sfermions, charginos, neutralinos and Higgs scalars in the
internal lines
induce a finite renormalization to the tree--level couplings of the
scalar potential, which may lead to a phase shift between the vacuum
expectation values of the two
neutral Higgs fields, {\it i.e.} to a spontaneous breaking of CP in the
scalar sector \cite{Maekawa,Pomarol}.

Unfortunately it comes out that this scenario is not realistic.
The point is that spontaneous CP violation can be implemented
radiatively only if a pseudoscalar with zero
tree--level mass exists, as was shown on general grounds by Georgi
and Pais \cite{Georgi}. In the MSSM this implies the existence of a very
light Higgs ( with a mass of a few GeV, given by one--loop contributions)
\cite{Pomarol}, which has been excluded by LEP data
\cite{Aleph}.

In the present paper we analyze the possibility of the spontaneous breakdown
of CP  at finite temperature in the MSSM. We find that new CP--violating
contributions arise from the one--loop corrections at $T \neq 0$, so
that the effective potential (at $T \neq 0$) may have a CP--non conserving
vacuum, while the pseudoscalar mass, calculated at $T=0$, is still compatible
with the experimental bounds. As $T$ goes to zero the phase of the
vacuum expectation values of the Higgs fields vanishes,
and the only
sources of CP violation remaining are the phase in the
Cabibbo-Kobayashi-Maskawa
matrix and the $\bar{\theta}$ parameter of the QCD vacuum.
Nevertheless, this effect may be of great relevance for the electroweak
baryogenesis in the MSSM, as it could give rise to a new, time varying,
CP--violating phase in
the scalar sector, whose size is nearly unbounded by the phenomenology at zero
temperature \footnote{See ref. \cite{Nelson} for a discussion on  the
relationship between baryogenesis and explicit CP violation
phenomenology in the MSSM.}.

The most general gauge invariant scalar potential for the
two--doublets model, along the neutral components, is given by
\begin{eqnarray}
V &=& {m_{1}}^{2}|H_1|^2 + {m_{2}}^{2}|H_2|^2 - ({m_{3}}^{2} H_1 H_2 + h.c.)
+\lambda_1 |H_1|^4 +\lambda_2 |H_2|^4 + \lambda_3 |H_1|^2 |H_2|^2 \nonumber\\
&+&\lambda_4|H_1 H_2|^2
+(\lambda_5 (H_1 H_2)^2  + \lambda_6 |H_1|^2 H_1 H_2 +\lambda_7 |H_2|^2
H_1 H_2 + h.c.),
\end{eqnarray}
where we assume ${m_3}^2$, $\lambda_5$, $\lambda_6$ and $\lambda_7$ to be real,
so that CP invariance may be broken only if the vacuum expectation
values of the Higgs fields get a non trivial phase,
\begin{equation}\delta\neq 0, \: \pi,
\end{equation} where we have defined $\langle H_1\rangle=v_1$,
$\langle H_2\rangle=v_2 \: e^{i\delta}$.

Eq. (2) is satisfied if, and only if,
\begin{equation}
\lambda_5>0,
\end{equation}
and
\begin{equation}
-1< \cos \delta = \frac{{m_3}^2-\lambda_6 {v_1}^2 -\lambda_7 {v_2}^2}
{4 \lambda_5 v_1 v_2} < 1.
\end{equation}

At the tree--level the parameters $\lambda_{i}$ $(i=1\dots 7)$ are fixed
by supersymmetry:
\begin{eqnarray}
\lambda_1 &=& \lambda_2\:\:=\:\:\frac{1}{4}({g_2}^{2}+{g_1}^{2}),\nonumber\\
\lambda_3 &=& \frac{1}{4}({g_2}^{2}-{g_1}^{2}),\nonumber\\
\lambda_4 &=& -\frac{1}{2}{g_2}^2, \nonumber \\
\lambda_5 &=& \lambda_6\:\:=\:\:\lambda_7\:\:=\:\:0,
\end{eqnarray}
where $g_2$ and $g_1$ are the gauge couplings of $SU(2)_L$ and $U(1)_Y$
respectively.

{}From eqs. (3) and (5) we immediately read that, at the tree--level,
CP is not violated in the scalar sector of the MSSM. Since it is the
soft breaking of supersymmetry that allows CP violation, the one--loop
contributions to the CP--violating
couplings $\lambda_5$, $\lambda_6$ and $\lambda_7$ will be proportional
to the soft parameters: the gaugino masses $M_{1,2}$, the sfermion
masses $\tilde{m}_F^2$, the trilinear scalar coupling $A$, and the
bilinear one, $B$. For this reason, we will include in the one--loop
effective potential only those field--dependent mass--matrices which
contain these parameters. The dominant contributions are given by the
stop for the bosonic sector and by chargino and neutralino for the
fermionic one.

At finite temperature the one--loop contribution to the effective
potential can be decomposed into the sum of a $T=0$ and a $T\neq0$ term:
\begin{eqnarray}
\Delta V_{T=0} &=& \frac{1}{64 \pi^2} {\rm Str} \left\{ {\cal M}^4
\left(\ln\frac{{\cal M}^2}{{\cal Q}^2}-\frac{3}{2}\right)\right\},\\
\Delta V_{T\neq 0} &=& \Delta V^{bos.}_{T\neq 0} + \Delta
V^{ferm.}_{T\neq 0}.
\end{eqnarray}
Defining $a^2_{b,(f)}\equiv{\cal M}_{b,(f)}^2/T^2$, where
${\cal M}_{b,(f)}$ is the bosonic (fermionic) mass matrix, the $T\neq 0$
contributions may be written  as

\begin {eqnarray}
\Delta V_{T \neq 0}^{bos.} &=& T^4\:\: {\rm Tr}' \left[ \frac{1}{24} a_b^2-
\frac {1}{12 \pi^2}(a_b^2)^{3/2}  - \frac{1}{64 \pi^2}a_b^4
\ln\frac{a_b^2}{A_b} \right. \nonumber \\
&-& \left. \pi^{3/2}\:\: {\sum}_{l=1}^{\infty}
(-1)^l \:\:\frac{\zeta(2l+1)}{(l+1)!}\:\:
\Gamma\left(l+\frac{1}{2}\right)\:\:
\left(\frac{a_b^2}{4 \pi^2}\right)^{l+2} \right],
\end{eqnarray}
\begin{flushright}
($a_b < 2 \pi$)
\end{flushright}
\begin{eqnarray}
\Delta V_{T \neq 0}^{ferm.}&=& T^4\:\: {\rm Tr}' \left[ \frac{1}{48}a_f^2 +
\frac{1}{64 \pi^2}a_f^4 \ln\frac{a_f^2}{A_f}\nonumber\right. \\
&+& \left. \frac{\pi^{3/2}}{8}\:\:
\sum_{l=1}^{\infty} (-1)^l\:\: \frac{1-2^{-2l-1}}{(l+1)!}\:\:
\zeta(2l+1)
\:\:\Gamma\left(l+\frac{1}{2}\right) \left(\frac{a_f^2}{\pi^2}\right)^{l+2}
\right],
\end{eqnarray}
\begin{flushright}
($a_f<\pi$)
\end{flushright}
\vskip 0.2 cm
where $A_b = 16 A_f = 16 \pi^2{\rm exp}(3/2 - 2 \gamma_E)$, $\gamma_E =
0.5772$, $\zeta$ is the Riemann $\zeta$-function, and ${\rm Tr}'$ properly
counts the degrees of freedom.
Eqs. (8) and (9) give an exact representation of the complete one--loop
effective potential at finite temperature \cite{Dolan} for $a_b< 2\pi$
and $a_f<\pi$, respectively.

Taking $\tilde{m}_Q^2=\tilde{m}_U^2$ the stop mass matrix takes the
convenient form
\begin{equation}
a_t^2=a_Q^2\cdot {\bf 1}  + \tilde{a}_t^2
\end{equation}
where $a_Q^2\equiv \tilde{m}_Q^2/T^2$, ${\bf 1}$ is the identity matrix,
and $\tilde{a}_t^2$ is the field--dependent part of the mass matrix, {\it
i.e.} $\tilde{a}_t^2\rightarrow 0$ as the fields vanish.
The only contributions to $m_3^2$, $\lambda_5$,
$\lambda_6$ and $\lambda_7$ come from the traces of $\tilde{a}_t^4$,
$\tilde{a}_t^6$ and $\tilde{a}_t^8$; the traces of the higher powers of
$\tilde{a}_t^2$ contain operators of dimension $d > 4$, which are
suppressed by powers of $H^2/\tilde{m}_Q^2$ or $H^2/(4 \pi^2 T^2)$,
multiplied by additional suppressing coefficient coming from the
expansion. Inserting eq. (10) in eq. (8) and
using the binomial expansion for the terms $\left(a_t^2\right)^{l+2}$
and $\left(a_t^2\right)^{3/2}$ we can extract the relevant terms from
the one--loop effective potential:
\begin{equation}
T^4\:\:\left[ -\frac{1}{32 \pi a_Q}- \frac{1}{64 \pi^2}
\left(\ln\frac{{\cal Q}^2}{A_b T^2} +\frac{3}{2}\right) -{\cal B}_4[a_Q^2]
\right] \:\:{\rm Tr}' \tilde{a}_t^4,
\end{equation}
\begin{equation}
T^4\:\: \left[ \frac{1}{192 \pi a_Q^3} -{\cal B}_6[a_Q^2] \right]
\:\:{\rm Tr}' \tilde{a}_t^6,
\end{equation}
\begin{equation}
T^4\:\: \left[ \frac{-1}{512 \pi a_Q^5} -{\cal B}_8[a_Q^2] \right]
\:\:{\rm Tr}' \tilde{a}_t^8,
\end{equation}
where,
\begin{equation}
{\cal B}_{2n}[a_Q^2]\equiv \pi^{3/2} \sum_{l={\rm max}[1,n-2]}^{\infty}\:\:
(-1)^l\:\:
\frac{\zeta(2l+1)}{(l+2)!}\:\:\Gamma\left(l+\frac{1}{2}\right) {l+2 \choose n}
\frac{(a_Q^2)^{l+2-n}}{(4 \pi^2)^{l+2}}
\end{equation}
\begin{flushright}
$n=2, 3, 4\ldots\:\:\:\:\:\:\:\:\:a_Q<2 \pi$.
\end{flushright}
In eq. (11) we have also included the contribution coming from the $T=0$
one--loop effective potential, eq.(6).
The numerical values of the series ${\cal B}_{2n}$
for some values of $a_Q$ are listed in table 1.

We now evaluate the traces in eqs. (11), (12) and (13) and find the
one--loop contribution to $m_3^2$ and to $\lambda_{5,6,7}$ coming from
the stop:
\beqra
\Delta {m_3^{(s)}}^2 &=& + 3 h_t^2 A_t T a_\mu \left[ \frac{1}{8 \pi a_Q}
+ \frac{1}{16 \pi^2}
\left(\ln \frac{{\cal Q}^2}{A_b T^2} + \frac{3}{2}\right) +
4 {\cal B}_4 [a_Q^2] \right],\\
 & &\nonumber \\
\Delta \lambda_5^{(s)}&=&-12 h_t^4 \frac{A_t^2 a_\mu^2}{T^2}
\left[ {\cal B}_8[a_Q^2] + \frac{1}{256 \pi a_Q^5} \right],\\
& & \nonumber \\
\Delta \lambda_6^{(s)}&=& - 6 h_t^2 \frac{A_t a_\mu}{T}
\left[ \frac{3}{4} (g_2^2+g_1^2)\left({\cal B}_6[a_Q^2]-\frac{1}{192 \pi
a_Q^3}\right)\right. \nonumber \\
&+&\left. 4 h_t^2 a_\mu^2 \left( {\cal B}_8[a_Q^2] +
\frac{1}{512 \pi a_Q^5}\right)\right],\\
& & \nonumber \\
\Delta \lambda_7^{(s)}&=& - 6 h_t^2 \frac{A_t a_\mu}{T}
\left[ \left(6 h_t^2 -\frac{3}{4}(g_2^2+g_1^2)\right)
\left({\cal B}_6[a_Q^2]-\frac{1}{192 \pi a_Q^3} \right)\right. \nonumber
\\
&+&\left. 4 h_t^2 \frac{A_t^2}{T^2} \left( {\cal B}_8[a_Q^2] +
\frac{1}{512 \pi
a_Q^5}\right)\right],
\eeqra
where we have defined $a_\mu\equiv \mu/T$.

Choosing $\mu^2 = M_1^2 = M_2^2$, the squared mass matrices for charginos
and neutralinos take a form analogous to that in eq. (10), i.e. they are
given by the sum of a multiple of the identity matrix and a field--dependent
matrix
\[
a_f^2=a_\mu^2 \cdot {\bf 1} + \tilde{a}_f^2\:\:.
\]
 We now insert them in eq. (9) and, following the same
strategy we used for the stop, we extract the relevant terms in the
one--loop potential,
\beq
T^4\:\:\left[ \frac{1}{64 \pi^2} \left(\ln\frac{{\cal Q}^2}{A_f T^2} +
\frac{3}{2}\right)+ {\cal F}_4[a_{\mu}^2]\right]\:\:{\rm Tr}'\tilde{a}_f^4,
\eeq
\beq
T^4\:\:{\cal F}_6[a_{\mu}^2]\:\:{\rm Tr}'\tilde{a}_f^6,
\eeq
\beq
T^4\:\:{\cal F}_8[a_{\mu}^2]\:\:{\rm Tr}'\tilde{a}_f^8,
\eeq
where the values of
\beq
{\cal F}_{2n}[a_{\mu}^2]=\frac{\pi^{3/2}}{8}
\sum_{l={\rm max}[1,n-2]}^\infty\:\: (-1)^l\:\: \frac{1-2^{-2l-1}}{(l+2)!}
\:\:\zeta(2l+1)\:\: \Gamma\left(l+\frac{1}{2}\right)
{l+2 \choose n} \:\:\frac{(a_{\mu}^2)^{l+2-n}}{(\pi^2)^{l+2}}
\eeq
are listed in table 2.
Finally, evaluating the traces in eqs. (19), (20) and (21), we obtain the
contributions to the relevant couplings from charginos,
\beqra
\Delta {m_3^{(c)}}^2&=& + {\rm sign}(\mu)\:\:  g_2^2 \: T^2\: a_\mu^2
\left[ \frac{1}{8 \pi^2}
\left(\ln \frac{{\cal Q}^2}{A_f T^2} +\frac{3}{2}\right) +
8 {\cal F}_4[a_{\mu}^2] \right]\\
\Delta \lambda_5^{(c)}&=& 8 \:g_2^4
a_\mu^4 {\cal F}_8[a_{\mu}^2],\\
\Delta \lambda_6^{(c)}&=&\Delta \lambda_7^{(c)}\:\:=\:\:
- {\rm sign}(\mu)\:\: 4\:\: g_2^4 \:\: a_\mu^2 \left[ 3 {\cal F}_6[a_\mu^2] +
4 a_\mu^2 {\cal F}_8[a_\mu^2]\right],
\eeqra
and from neutralinos,
\beq
\Delta {m_3^{(n)}}^2 = \frac{(g_2^2 +g_1^2)}{2 g_2^2} \Delta
{m_3^{(c)}}^2,\:\:\:\:\:\:\:\:\:\:\:
\Delta \lambda_i^{(n)} = \frac{(g_2^2 +g_1^2)}{2 g_2^2}\Delta
\lambda_i^{(c)},\:\:\:\:\:i=5,6,7.
\eeq

Now we are ready to discuss the conditions (3) and (4) on the spontaneous
CP breaking at finite temperature and their implication on the mass
spectrum of the Higgs scalars. The one--loop effective potential at
finite temperature has a  CP--violating minimum if (see eqs. (3) and (4))
\beq \Delta \lambda_5 >0,
\eeq
and
\beq
(1-{\rm K})<\frac{\bar{m}_3^2}{\Delta \lambda_6 v_1^2(T) +
\Delta \lambda_7 v_2^2(T)}<(1+{\rm K})
\eeq
where
\[ {\rm K}\equiv 4 \frac{\Delta \lambda_5}{\Delta \lambda_6 +\Delta
\lambda_7 \tan^2 \beta (T)} \tan \beta (T),
\]
$v_{1,2}(T)$  are the vacuum
expectation values at finite temperature, $\tan \beta(T)\equiv v_2 (T)/v_1
(T)$, $\Delta \lambda_i\equiv
\Delta \lambda_i^{(s)}+\Delta
\lambda_i^{(c)}+\Delta \lambda_i^{(n)}$ ($i= 5, 6, 7$),  and $\bar{m}_3^2$
is the coefficient of the operator $- H_1 H_2$ in the one--loop effective
potential at $T\neq0$, that is
\beq
\bar{m}_3^2 \equiv m_3^2 + \Delta {m_3^{(c)}}^2 +\Delta {m_3^{(n)}}^2
+\Delta {m_3^{(s)}}^2.
\eeq
The radiatively-corrected mass of the pseudoscalar is given
by
\beqra
m_A^2&=&\frac{1+\tan^2 \beta }{\tan \beta}\left[ m_3^2 - {\rm sign}(\mu)
\frac{\mu^2}{16 \pi^2} (3 g_2^2 + g_1^2) \ln\frac{\mu^2}{{\cal
Q}^2}-\frac{3}{16 \pi^2} h_t^2 A_t \mu \ln \frac{\tilde{m}_Q^2}{{\cal
Q}^2}\right]\nonumber \\
&=&\left.\frac{1+\tan^2 \beta }{\tan \beta} \right\{ \bar{m}_3^2 -
{\rm sign}(\mu)\:\: (3 g_2^2 + g_1^2)\: T^2\: a_{\mu}^2 \left[
\frac{1}{16 \pi^2}\left(\ln\frac{a_{\mu}^2}{A_f}+\frac{3}{2}\right)
+ 4 {\cal F}_4 [a_\mu^2 ]\right]  \nonumber \\
 &-&\left. 3 \: h_t^2\: A_t \:T\:a_\mu \left[ \frac{1}{8 \pi a_Q} +
\frac{1}{16 \pi^2}\left(\ln \frac{a_Q^2}{A_b} +\frac{3}{2}\right) + 4
{\cal B}_4 [a_Q^2 ]\right]\right\},
\eeqra
where we have eliminated $m_3^2$ using eq. (29).
{}From eq. (30) we read that in the limit \hbox{$g_1$, $g_2$, $h_t \rightarrow
0$} the pseudoscalar mass vanishes if we require spontaneous CP violation
(eq. (28)), so, in agreement with Georgi-Pais theorem \cite{Georgi},
$m_A^2$ is a
one--loop
effect.
Nevertheless, it does not
imply a very light pseudoscalar, as in the case of spontaneous
CP violation at $T=0$ \cite{Pomarol}. In fact, the contributions at
$T\neq0$ may be important, as we can read off from eq.(30): even if
$\bar{m}_3^2$ is constrained to be of the same order of $\Delta \lambda_6 v_1^2
(T)+ \Delta \lambda_7 v_2^2 (T)$, as required by spontaneous
CP violation, eq. (28), the second and the
third terms in the R.H.S. of the last line in eq.(30) may give significant
contributions to $m_A^2$.

The present experimental lower bound on $m_A^2$ comes from LEP
\cite{Aleph}; it is about 20 GeV for $\tan \beta=1$ and
rapidly saturates the kinematical limit for LEP search, $m_A^2 > 40$
GeV, as $\tan \beta$ becomes greater than about 1.5.
In Fig. 1 we plot the region in the plane ($\mu$, $\tilde{m}_Q$) in which
$m_A (\tan \beta/ (1 +\tan^2 \beta))^{1/2}$ is greater than 14.1 GeV and
17.7 GeV, that is,
$m_A > 40$ and 50 GeV, respectively, for $\tan \beta =4$.
We have fixed $\bar{m}_3^2=\Delta \lambda_6 v_1^2 (T)
+ \Delta \lambda_7 v_2^2 (T)$, which corresponds to the
maximal value for the CP--violating phase (see eq. (4)), $\cos \delta =
0$, and also required that $\Delta \lambda_5 >0$.
As we can see there is a wide region in the parameters space in which CP
is broken and $m_A$ is compatible with the experimental values.
Moreover, we want to stress that our choice $\mu^2=M_2^2=M_1^2$ and
$\tilde{m}_U^2=\tilde{m}_Q^2$ has been made only to simplify the
derivation in an effective potential approach, but
the CP violation is present in a much larger portion of the parameters
space.

Since we believe that this effect may play a crucial role in the electroweak
baryogenesis, we have fixed $T=T_{e.w.}=O(150 {\rm GeV})$ \cite{Giudice}
in fig. 1. As $T$ decreases the values of $v_{1,2}(T)$, $\bar{m}_3^2$,
$\Delta \lambda_i$, and consequently $\cos \delta$, change, and at a
certain value $T=T_{rest.}$, the condition (28) is no more fulfilled,
{i.e.} CP is restored. It is worthwhile to mention that in this formalism
we are not allowed to take the $T \rightarrow 0$ limit in the effective
potential, because our formulas (8) and (9) are valid only for
$T>\tilde{m}_Q/(2\pi)$, $\mu/\pi$.

Another possibility is that the vacuum expectation values $v_{1,2}
(T_{e.w.})$ in the middle of the electroweak bubbles are too large, so
that the condition (28) is not fulfilled (this might follow from the
requirement that the anomalous B-- and L--violating processes are out of
equilibrium \cite{shaposhnikov}); in this case CP violation can take place
in the bubble walls, where the vacuum expectation values are changing
from zero to the value inside the bubble.

In order to make more quantitative statements on the role of this effect
in the generation of the baryon asymmetry of the Universe, a detailed
analysis of the phase transition and of bubble
propagation in the MSSM is needed. It will be the subject of a
forthcoming publication \cite{comelli}.

In conclusion, we have shown that the spontaneous CP violation is possible
in the Minimal Supersymmetric Standard Model at finite temperature, and
still in agreement with the experimental lower bounds on the mass of the
Higgs pseudoscalar. The CP breaking may take place soon after the
electroweak phase transition, and the CP--violating phase may reach the
maximal values $\delta=\pm \pi/2$, going to zero as the temperature of the
Universe falls down.
\\
\begin{center}
{\bf Acknowledgments.}
\end{center}
It is a pleasure to thank A. Brignole, G. R. Dvali, G. Giudice,
A. Riotto and G. Senjanovi\'{c} for useful discussions,
and A. Masiero and C. Verzegnassi
for the continuous encouragement. We are grateful also to M. D'Attanasio,
R. De Pietri and G. Marchesini for the kind hospitality at the department
of physics of Parma, where part of this work was done.
\newpage
%
\def\MPL #1 #2 #3 {Mod.~Phys.~Lett.~{\bf#1}\ (#3) #2}
\def\NPB #1 #2 #3 {Nucl.~Phys.~{\bf#1}\ (#3) #2}
\def\PLB #1 #2 #3 {Phys.~Lett.~{\bf#1}\ (#3) #2}
\def\PR #1 #2 #3 {Phys.~Rep.~{\bf#1}\ (#3) #2}
\def\PRD #1 #2 #3 {Phys.~Rev.~{\bf#1}\ (#3) #2}
\def\PRL #1 #2 #3 {Phys.~Rev.~Lett.~{\bf#1}\ (#3) #2}
\def\RMP #1 #2 #3 {Rev.~Mod.~Phys.~{\bf#1}\ (#3) #2}
\def\ZP #1 #2 #3 {Z.~Phys.~{\bf#1}\ (#3) #2}

\newpage

\noindent{\bf Figure Caption}
\begin{itemize}
\item[{\bf Fig. 1}]{The parameters space (above the dashed line) in the plane
$(\mu,\tilde{m}_Q)$ for $\cos \delta = 0$, $\lambda_5>0$ (see text).
Contours corresponding to
$m_A (\tan \beta/ (1+\tan^2 \beta))^{1/2} >14.1$ and $17.7 {\rm GeV}$
are plotted.
We have fixed $A_t=50 {\rm GeV}$, $T=150 {\rm GeV}$, $v_1(T)=v_2(T)=90
{\rm GeV}$ $(v(T=0)=174 {\rm GeV})$ and $h_t=1$.}
\end{itemize}

\vskip 2.cm
\begin{itemize}
\item[{\bf Tab. 1}]{Values of the series ${\cal B}_{2n}[a_Q^2]$ for values of
$a_Q^2$ in the allowed range.}
\end{itemize}
\begin{tabular}{|c||c||c||c||}\hline
$a_Q$  &  ${\cal B}_4[a_Q^2]$ & ${\cal B}_6[a_Q^2]$ & ${\cal B}_8[a_Q^2]$
\\    \hline
  1 &  $-4.74 \cdot 10^{-5}$ & $-1.56 \cdot 10^{-5}$ & $1.24\cdot 10^{-7}$ \\
  2 &  $-1.81\cdot 10^{-4}$ & $-1.42\cdot 10^{-5}$ & $1.04 \cdot10^{-7}$ \\
  3 &  $-3.8 \cdot 10^{-4}$ & $-1.24\cdot 10^{-5}$ & $8.01 \cdot10^{-8}$ \\
  4 &  $-6.18\cdot 10^{-4}$ & $-1.05\cdot10^{-5}$ & $5.8\cdot  10^{-8}$ \\
  5 &  $-8.76\cdot 10^{-4}$ & $-8.71\cdot 10^{-6}$ & $4.06\cdot 10^{-8}$ \\
  6 &  $-1.14 \cdot10^{-3}$ & $-7.22\cdot 10^{-6}$ & $2.81\cdot 10^{-8}$ \\
\hline
\end{tabular}
\vskip 2. cm
\begin{itemize}
\item[{\bf Tab. 2}] {Values of the series ${\cal F}_{2n}[a_\mu^2]$ for
values of
$a_\mu^2$ in the allowed range.}
\end{itemize}

\begin{tabular}{|c||c||c||c||}\hline
$a_\mu$  &  ${\cal F}_4[a_\mu^2]$ & ${\cal F}_6[a_\mu^2]$ & ${\cal
F}_8[a_\mu^2]$
\\    \hline
  0.5 &  $-8.29\cdot 10^{-5}$ & $-1.09 \cdot10^{-4}$ & $3.84\cdot 10^{-6}$ \\
  1   &  $-3.15 \cdot10^{-4}$ & $-9.8 \cdot 10^{-5}$ & $3.21\cdot 10^{-6}$ \\
  1.5 &  $-6.55\cdot 10^{-4}$ & $-8.4 \cdot 10^{-5}$ & $2.45\cdot 10^{-6}$ \\
   2  &  $-1.06 \cdot10^{-3}$ & $-6.95\cdot 10^{-5}$ & $1.75\cdot 10^{-6}$ \\
  2.5 &  $-1.48\cdot 10^{-3}$ & $-5.63\cdot 10^{-5}$ & $1.21\cdot 10^{-6}$ \\
   3  &  $-1.89\cdot 10^{-3}$ & $-4.54\cdot 10^{-5}$ & $8.18\cdot 10^{-7}$ \\
\hline
\end{tabular}

\end{document}